\pdfoutput=1
\documentclass[aps,prl,nofootinbib,preprintnumbers,amsmath,amssymb,latexsym,array,enumerate,letter,twocolumn]{revtex4}
\usepackage{mathtools}
\usepackage{epsfig}
\usepackage{graphicx}
\usepackage{color}
\usepackage{amssymb}
\usepackage{enumitem}
\usepackage{amsmath}
\usepackage{hyperref}
\usepackage{breakurl}

\def\meff{m_h}
\def\Teff{T_{\text{eff}}}

\makeatletter

\def\be{\begin{equation}}
\def\ee{\end{equation}}
\newcommand{\bea}{\begin{eqnarray}}
\newcommand{\eea}{\end{eqnarray}}

\begin{document}

%\widetext
\preprint{DESY-23-004}

\title{The Tachyonic Higgs and the Inflationary Universe}

\author{Bibhushan Shakya}

\affiliation{Deutsches Elektronen-Synchrotron DESY,
Notkestr.\,85, 22607 Hamburg, Germany}

\begin{abstract}

The Standard Model Higgs becomes tachyonic at high energy scales according to current measurements. This unstable regime of the Higgs potential can be realized in the early Universe during high scale inflation, potentially with catastrophic consequences. This letter highlights a crucial inherent feature of such configurations that has so far remained ignored: Higgs particle production out of vacuum induced by the rapidly evolving Higgs field, which gets exponentially enhanced due to the tachyonic instability. Such explosive particle production can rapidly drain energy away from the Higgs field, sustaining a significant density of Higgs particles even during inflation, and could initiate a qualitatively different form of preheating in parts of the post-inflationary Universe. Any study of the Higgs field in its tachyonic phase, either during or after inflation, must therefore take this substantial particle energy density into account, which could significantly affect the subsequent evolution of such systems. This could carry important implications for high scale inflation, post-inflationary preheating, observable signals in the cosmic microwave background, gravitational waves, and primordial black holes, as well as deeper concepts ranging from eternal inflation to the metastability of the electroweak vacuum. 
  
\end{abstract}

\maketitle

\section{Motivation}

Current measurements indicate that the Standard Model (SM) Higgs potential is unstable at high scales, and the electroweak  (EW) vacuum that our Universe exists in is metastable, albeit with a decay lifetime significantly longer than the current age of the Universe. However, the Higgs could have briefly existed in this unstable regime in the early Universe due to quantum fluctuations during a period of high scale inflation. Such configurations have been extensively studied in the literature \cite{Ellis:2009tp,Espinosa:2007qp,Elias-Miro:2011sqh,Kobakhidze:2013tn,Hook:2014uia,Kobakhidze:2014xda,Shkerin:2015exa,Kearney:2015vba,Espinosa:2015qea,East:2016anr,Grobov:2016llk,Kohri:2016qqv,Markkanen:2018pdo,DeLuca:2022cus}, and the consequences are believed to be catastrophic: the Higgs field rapidly evolves to regions of negative potential energy that can terminate inflation, resulting in crunching anti-de Sitter (AdS) space that grows to engulf all of spacetime, rendering the existence of a Universe such as ours impossible. This fate can be avoided with nonminimal modifications of the Higgs potential that stabilize it before reaching such regimes (see e.g.\,\cite{Lebedev:2012sy,Fairbairn:2014zia,Herranen:2014cua,Kamada:2014ufa,Kearney:2015vba,Espinosa:2015qea,Saha:2016ozn,Enqvist:2016mqj,Ema:2017ckf,Ema:2017loe,Figueroa:2017slm,Postma:2017hbk,Fumagalli:2019ohr,Kost:2021rbi}). However, in the absence of such stabilizing corrections, the SM Higgs appears to be incompatible with inflation scales greater than the instability scale of the Higgs potential.

In this Letter, we study the effects of Higgs particle production in the tachyonic regime during inflation. It is well known that the tachyonic instability triggers an expontential growth of particle number \cite{Felder:2000hj,Felder:2001kt,Boyanovsky:1996sq}. Some previous papers \cite{RodriguezRoman:2018tri,Espinosa:2017sgp,DeLuca:2022cus} that considered particle production and tachyonic growth of inhomogeneities in this regime during inflation found such effects to be negligible; however, these papers only considered Hubble-induced effects,\,\textit{i.e.}\,those sourced by the inflationary background. In this paper, we focus on particle production induced by \textit{the dynamics of the Higgs field itself}. It is well known that a non-adiabatically changing background field can produce particles out of vacuum; this phenomenon is encountered in many familiar contexts, such as the Schwinger mechanism, Hawking radiation from black holes, or gravitational particle production. Although the energy density in the Higgs field is subdominant to the inflaton energy density in our regime of interest, which might have led previous studies to ignore this effect, we will see that particle production induced by the Higgs field evolution is an important effect, due to the fact that the Higgs field can reach significantly larger values than the Hubble scale during inflation. 

A substantial population of Higgs particles produced out of the Higgs field during inflation can have several important consequences. It can draw energy out of the Higgs field, slowing its evolution towards catastrophic values, as well as produce stabilizing thermal corrections to the Higgs potential. It can terminate inflation locally once its energy density becomes comparable to the inflaton energy density, resulting in emergence out of inflation into a preheated state, much as in warm inflation scenarios \cite{Rangarajan:2018tte}, rather than into AdS. Even the collapse into AdS, currently believed to be catastrophic, could become benign due to modified evolution due to the significant energy density in particles. Such considerations reopen the possibility of restoring the Universe to the EW vacuum after reheating, thereby making high scale inflation  compatible with the Higgs instability.  The presence of a large density of particles in some Hubble patches could also lead to various observables signals of such inhomogeneities, such as imprints in the cosmic microwave background (CMB), gravitational waves, and primordial black holes.

The main purpose of this Letter is to demonstrate that excursions of the Higgs to large field values of its unstable potential is necessarily accompanied by a huge energy density of Higgs particles, even during inflation, which can affect the subsequent evolution of the Higgs field.  Section \ref{sec:higgs} describes the framework for the study. Section \ref{sec:particleproduction} presents the calculation of particle production from Higgs evolution and tachyonic instability during inflation. Backreaction effects of particle production are addressed in Section \ref{sec:backreaction}, followed by qualitative discussions of the post-inflationary evolution of such regions (Section \ref{sec:postinflation}) and observable signals of such configurations (Section \ref{sec:observables}).  Section \ref{sec:discussion} is devoted to a discussion of open questions and broader implications.

\section{Framework: Higgs Evolution}
\label{sec:higgs}

The Standard Model Higgs potential develops an instability scale at $\Lambda_I\sim 10^{11}$ GeV due to the Higgs quartic coupling running to negative values (see e.g.\,\cite{Buttazzo:2013uya}). Above this scale, the Higgs potential can be written as 
\be
V(h)\approx-\frac{\lambda}{4}h^4\,.
\label{eq:higgspotential}
\ee
We can approximate $\lambda\approx 0.01$ for our purposes. The (field-dependent) Higgs mass in this regime is tachyonic:
\be
\meff^2(h)=V_{hh}=-3\lambda h^2\approx-(0.17\, h)^2<0\,.
\label{eq:meff}
\ee

The inflaton potential is
\be
V_{\phi}=\frac{3}{8\pi}H^2 M_P^2\,,
\ee
where $M_P\approx 1.2\times 10^{19}\,$GeV is the Planck scale, and $H$ is the Hubble scale during inflation, which we take to be constant. 

The evolution of the Higgs field in this setup has been studied in detail in several previous works \cite{Espinosa:2007qp,Kobakhidze:2013tn,Hook:2014uia,Kearney:2015vba,Espinosa:2015qea,East:2016anr,Markkanen:2018pdo,DeLuca:2022cus}. For small (sub-Hubble) Higgs field values, the dynamics is dominated by quantum fluctuations of size $\sim\frac{H}{2\pi}$ induced by inflation, resulting in random coherent ``jumps" of the Higgs field within entire Hubble patches, which remains the dominant driving force until the Higgs reaches $h\approx (3/2\pi\lambda)^{1/3}H\approx 3.6 H$. Beyond this, classical evolution driven by the Higgs potential takes over, and the equation of motion of the Higgs field is
\be
\ddot{h}+3H\dot{h}=\frac{dV}{dh}\,.
\label{eq:higgseom}
\ee
We will solve for Higgs evolution in this classical regime, with initial conditions $h=3.6 H$ and $\dot{h}=0$, to obtain the Higgs field value as a function of time, $h(t)$. In the early stages of this regime, Hubble friction causes the Higgs field to slow-roll for several e-folds of inflation, until it reaches 
\be
h\sim h_{sr}\equiv (3/\lambda)^{1/2} H\approx 17.3 H~~~(\text{end of slow-roll})
\ee
Beyond this point, Hubble friction becomes negligible, and the Higgs field diverges quickly to very large values in less than a single e-fold. 

Note that the inflaton energy density dominates over the Higgs potential energy until  
\be
h\sim h_I\equiv\left(\frac{3}{2\pi \lambda}\right)^{1/4}\!\!\sqrt{H\,M_P}~~~(\text{exit from inflation})
\ee
For this paper, we take the scale of inflation to be of the same order as the Higgs instability scale, $H\sim\Lambda_I$; then the inflaton energy density dominates until $h\gtrsim h_I\sim10^4 H$. If the Higgs field gets to such large values in a Hubble patch, this terminates inflation locally, and the region rapidly collapses into AdS. These collapsing regions grow to engulf the surrounding spacetime after inflation has ended globally  \cite{East:2016anr}; therefore, the existence of even a single Hubble patch where the Higgs field extends beyond the slow-roll regime is posited to be catastrophic for the existence of our  Universe \cite{East:2016anr,Espinosa:2015qea,Strumia:2022kez}. However, as we will see below, it is precisely in this window beyond slow-roll, $h_{sr}< h<h_I$, that particle production becomes important, and could affect the subsequent evolution.

\section{Particle Production and Tachyonic Growth}
\label{sec:particleproduction}

We now consider particle production from the evolving Higgs field in the regime $h_{sr}< h<h_I$. It is well known that particle production during inflation requires non-adiabatic conditions. Beyond the Higgs slow-roll regime $h > h_{sr}$, the Higgs mass indeed changes non-adiabatically, $|\dot{m}_{h}/\meff^2|\sim 1$, as can be verified numerically using Eq.\,\ref{eq:meff} and the numerical solution for Eq.\,\ref{eq:higgseom}. This non-adiabatic evolution of the Higgs mass can therefore excite Higgs particles out of the vacuum.\,\footnote{All other SM particles that obtain mass from the Higgs mechanism also have non-adiabatically varying masses and can get excited out of vacuum in this phase; however, their masses are not tachyonic, hence the effects of their production are negligible.} The standard approach to calculate the number density of particles produced from a non-adiabatically changing background is via the computation of Bogoliubov coefficients (see e.g.\,\cite{Birrell:1982ix,Parker:2009uva,Gorbunov:2011zzc}). Here, we first present a semi-analytic estimate that is computationally simpler and offers greater intuition before comparing with numerical solutions. 

When the Higgs mass evolves non-adiabatically, modes with momenta $k\lesssim|\meff|$ get populated with occupation number $n_k=|\beta_k|^2\sim 1$, where $\beta_k$ is the Bogoliubov coefficient of a positive frequency mode, corresponding to particle excitation \cite{Birrell:1982ix,Parker:2009uva,Gorbunov:2011zzc,GarciaBellido:2001cb}. When the mass is tachyonic, the coefficient gets further enhanced exponentially via the tachyonic instability as $\beta\sim e^{-i\omega t}= e^{|\omega| t}$, where $\omega^2=\meff^2+k^2$, for modes with $\omega^2<0$. The Higgs particle energy density as a function of the Higgs field value can therefore be estimated as\,\footnote{Strictly speaking, the interpretation of $n_k$ as the number of particles with energy $|\omega_k|$ is robust only at a stable point of the theory, not in the unstable regime while the background is changing; nevertheless we will adopt this interpretation here, as is commonly done in the literature.}
\be
\rho_P(h)=\!\int_H^{\meff}\!\!\frac{d^3 k}{(2\pi)^3}|\omega(h,k)| n_k=\frac{1}{2\pi^2}\!\!\int_H^{\meff}\!k^2 dk |\omega(h,k)| n_k
\label{eq:density1}
\ee
where $\omega^2\,(h,k)=\meff^2(h)+k^2.$ The mode occupation number is evaluated as
\be
n_k=|\beta_k|^2\approx \text{Exp}{\left[\,2\int |\omega(h,k)| dt\right]}\, ,
\label{eq:number}
\ee
where the integral is taken over all tachyonic regimes, i.e.\,over all $k$ and $t$ where $\omega^2<0$ holds.

For particle production during inflation, two additional considerations must be taken into account:
\begin{itemize}
\item Inflation redshifts momenta and dilutes number densities exponentially fast: $k\to k/a\approx k e^{-Ht}$. 
\item The amplitudes of modes that become larger than the horizon size during inflation, ie $k\!<\!H$, get frozen, and do not grow further.
\end{itemize}
Thus, exponential growth can only occur for modes in the window $ H< k <|\meff|$, which sets the limits of the integral in Eq.\,\ref{eq:density1}. Since momenta are exponentially redshifted, it might appear that sub-horizon modes have no time to grow substantially before crossing the horizon. However, as mentioned earlier, the rapidly evolving Higgs field transitions from $h_{sr}$ to $h_I$ within a single e-fold of inflation, hence the above effects are not substantial (but are nevertheless incorporated in the calculation). The crucial point to note here is that even a fraction of an e-fold can realize a significant growth factor $e^{|\omega| t}\sim e^{|\meff| t}$ when $|\meff|\gg H$. 

The integral in the exponent in Eq.\,\ref{eq:number} at time $t$ can be evaluated as
\be
\int_{t_i}^{t} |\omega(h,k)| dt_v = \!\int_{t_i}^{t}\!\!\sqrt{\text{Max}[-\meff^2(t_v)-(k e^{(t-t_v)})^2,\,0]}\,dt_v\,.
\ee
Here, the Max function restricts the integral to the tachyonic regimes $\omega^2<0$, whereas the $e^{(t-t_v)}$ factor accounts for the exponential redshifting of momenta due to inflation. We take the starting point $t_i$ as the time at which $h\approx h_{sr}$, as this is when the Higgs evolution becomes non-adiabatic; in practice, the integral is dominated by contributions from later times, when $|\meff|$ is larger, and is therefore fairly insensitive to the exact choice of $t_i$.  

As a cross-check, the estimate in Eq.\,\ref{eq:number} can be compared with numerical solutions for the Bogoliubov coefficients. Ignoring the expansion of space, the mode function $X_k$ for a scalar field evolves as 
\be
X_k''+(k^2 +\meff^2(t))X_k=0\,.
\label{eq:bogoliubovmode}
\ee
The field can be set to be in vacuum at time $t_i$ by choosing the initial conditions $X_k(t_i)=1/\sqrt{2\omega_{k,0}}$, $X'_k(t_i)=-i\sqrt{\omega_{k,0}/2}$, where $\omega^2_{k,0}=k^2+\meff^2(t_i)$. The mode occupation number at subsequent times is then evaluated as
\be
n_k(t)=|\beta_k (t)|^2=\frac{|X'_k|^2+|\omega_k\,X_k|^2}{2|\omega_k|}-\frac{1}{2}\,.
\label{eq:bogoliubov}
\ee
Choosing $t_i$ as the time when $h=h_{sr}$ (again, the numerical results are also insensitive to the exact choice of $t_i$), it can be verified that the estimate for $|\beta_k|$ in Eq.\,\ref{eq:number} agrees with the numerical results from the above prescription (solving Eq.\,\ref{eq:bogoliubovmode},\,\ref{eq:bogoliubov})  within an $\mathcal{O}(1)$ factor for all values of $(k,h)$ of interest. With this confirmation, we will henceforth use Eq.\,\ref{eq:number}, which is computationally simpler, to calculate the occupation number of particles.

\begin{figure}[t]
\begin{center}
 \includegraphics[width=0.9\linewidth]{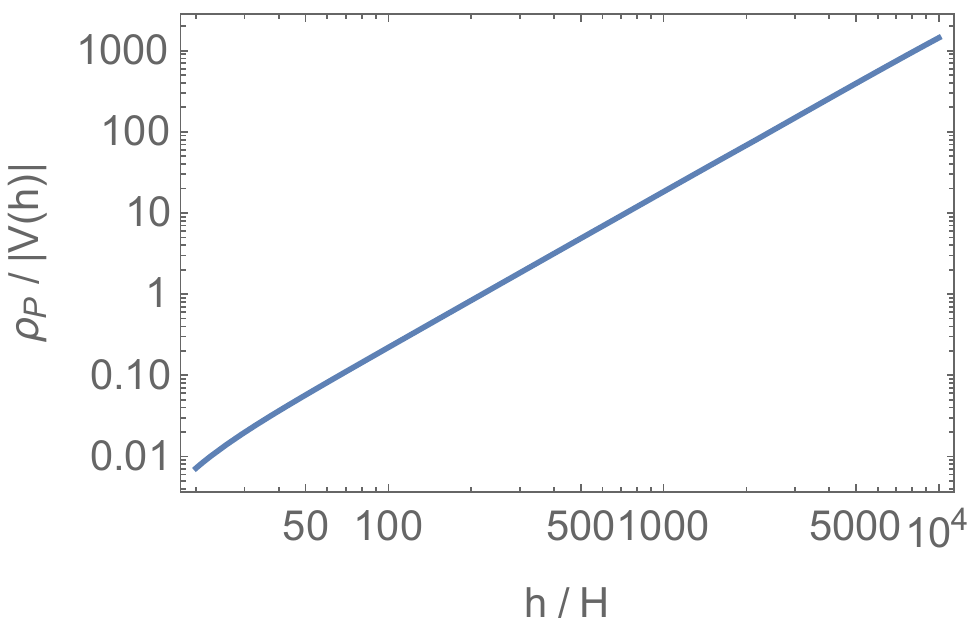}
  \end{center}\vspace{-4mm}
\caption{Ratio of energy density in Higgs particles to the potential energy of the Higgs field, as defined in Eq.\,\ref{eq:ratio}, without taking any backreaction effects into account.}
\label{fig:ratio}
\end{figure}

Next, consider the ratio of energy density in Higgs particles to the Higgs potential energy at a given time $t$
\be
\frac{\rho_P(t)}{|V(h(t))|}=\frac{2}{\pi^2\lambda}\int_H^{|\meff(t)|} \frac{k^2 dk \,|\omega(h,k)|\, n_k(t)}{h^4 (t)}\,.
\label{eq:ratio}
\ee
Fig.\,\ref{fig:ratio} plots this ratio in the regime $h_{sr}< h<h_I$. The particle energy density starts as a sub-percent level effect at $h_{sr}$, but becomes comparable to the potential energy in the Higgs field for $h\gtrsim 220\, H$, at which point backreaction effects of particle production need to be taken into account. Numerically, it is found that the integral is dominated by the lowest values of $k$, since these are the modes that have been growing exponentially for the longest duration.

The growth of this ratio can be understood as follows. Approximating $|\omega(h,k)|\approx |\meff|=\sqrt{3\lambda} h$, we have
\be
\frac{\rho_P(t)}{|V(h(t))|}=\frac{2\sqrt{3}}{\pi^2\sqrt{\lambda}}\int_H^{|\meff (t)|} \left(\frac{k \, |\beta_k|}{h}\right)^2 \frac{dk}{h}\,. 
\label{eq:ratio2}
\ee

The prefactor is $\mathcal{O}(1)$ and can be ignored. In the absence of exponential growth, $\beta_k\sim 1$, and the integral evaluates to $(1/3)|m_{h}^3|$, giving $\frac{\rho_P(t)}{|V(h(t))|}\sim (1/3) (3\lambda)^\frac{3}{2}\approx 10^{-3}$, a sub-percent level effect (as also seen in Fig.\,\ref{fig:ratio} for $h\sim h_{sr}$). However, the expression in Eq.\,\ref{eq:ratio2} suggests that if $k |\beta_k|$ grows faster than $h$ with time, an increasingly larger fraction of the Higgs potential energy can get transferred to Higgs particles. 

An insightful way to understand that this indeed occurs is to note that the Bogoliubov coefficient $|\beta_k|$ scales as the Higgs velocity $\dot{h}$ for small $k$. This can be seen clearly, as pointed out in \cite{Espinosa:2017sgp}, by taking the time derivative of the Higgs equation of motion Eq.\,\ref{eq:higgseom}, which reveals that $\dot{h}$ also follows the simple harmonic oscillator equation with negative frequency $-|\meff|$. Hence both $\dot{h}$ and $|\beta_k|$  grow exponentially as $\sim e^{|m_h| t}$. This observation enables us to write 
\be
|\beta_k (t)|\sim \frac{\dot{h} (t)}{\dot{h}_{h=h_{sr}}}\approx \frac{h^2}{h_{sr}^2}~~~~~(\text{for~} k\ll |\meff|)
\label{eq:nkapprox}
\ee
For the final step, we have used the fact that when Hubble friction is negligible, energy conservation dictates that the potential energy lost by the Higgs field gets converted entirely into its kinetic energy, $\frac{1}{2}\dot{h}^2\approx \frac{\lambda}{4}h^4$. With this, $k |\beta_k|/h\sim k h /h_{sr}^2$, which clearly grows rapidly with time. Note, in particular, that for the lowest modes $k\sim H$, $k |\beta_k|/h\sim H h /h_{sr}^2$ becomes $\mathcal{O}(1)$ for $h \sim (h_{sr}/H)^2 H \sim 300\,H$, which is consistent with Fig.\,\ref{fig:ratio}.
 
Let us briefly discuss aspects that may disrupt the exponential enhancement of Higgs number density.  The most worrisome is the decay of Higgs particles into other SM particles. Since the coherent Higgs field value $h$ sets the background vacuum expectation value (vev) in the Hubble patch, all SM particles are correspondingly heavier in this dynamic phase. In particular, since the Higgs mass is $|\meff|$, decays into SM states that interact most strongly with the Higgs, i.e.\,W and Z gauge bosons and the top quark, are kinematically blocked, and the dominant decay is into $b\bar{b}$ and $WW^{*}$. The decay rate into $b\bar{b}$, for instance, is $\Gamma_h\sim 3 y_b^2/(8 \pi ) |\meff|\sim 10^{-5} h$, thus $\Gamma_h<H$ in the regime of interest $h_{sr}< h<h_I$, and Higgs decays can be neglected. The rate for annihilation processes is parametrically similar, hence annihilations are likewise negligible.

\section{Backreaction Effects}
\label{sec:backreaction}

The previous section has demonstrated that the rapidly evolving tachyonic Higgs field triggers an exponential growth of particles, whose energy density naively exceeds that available from the Higgs potential for $h\gtrsim 220 H$. Hence it becomes necessary to take into account the backreaction of the population of particles on the evolution of the Higgs field. The production of particles affects Higgs dynamics in two ways. First, particle production takes energy out of the Higgs field, slowing its evolution towards higher field values. Second, it introduces thermal corrections to the Higgs potential. The former effect is more important, hence we address it first. 

The energy loss from the Higgs field due to particle production can be incorporated into the Higgs equation of motion Eq.\,\ref{eq:higgseom} by adding a new friction term to account for the energy transferred to particles. Its effect will be to slow the evolution of the Higgs, i.e.\,decrease $\dot{h}$, which will in turn suppress the production of particles (Eq.\,\ref{eq:nkapprox}). Since the energy density of particles itself involves integrating over the Higgs evolution up to that point (Eq.\,\ref{eq:number},\,\ref{eq:ratio}), this results in a nontrivial, coupled system that must be solved numerically. We will, instead, make use of analytical estimates that are sufficient for a qualitative understanding of the evolution of the system. 

The key insight is to make use of the observation that the particle mode occupation number scales as the square of the Higgs velocity, $n_k\sim \dot{h}^2$ (recall Eq.\,\ref{eq:nkapprox}). This can be combined with Eq.\,\ref{eq:ratio2} to obtain an approximate expression for the energy density in Higgs particles
\be
\rho_P(h)\sim\left(\frac{\lambda}{3}\right)^2\left(\frac{h}{H}\right)^2 \dot{h}^2\,.
\label{eq:rhoest}
\ee
This expression provides an excellent match to Fig.\,\ref{fig:ratio}, as can be checked numerically.

\begin{figure}[t]
\begin{center}
 \includegraphics[width=0.9\linewidth]{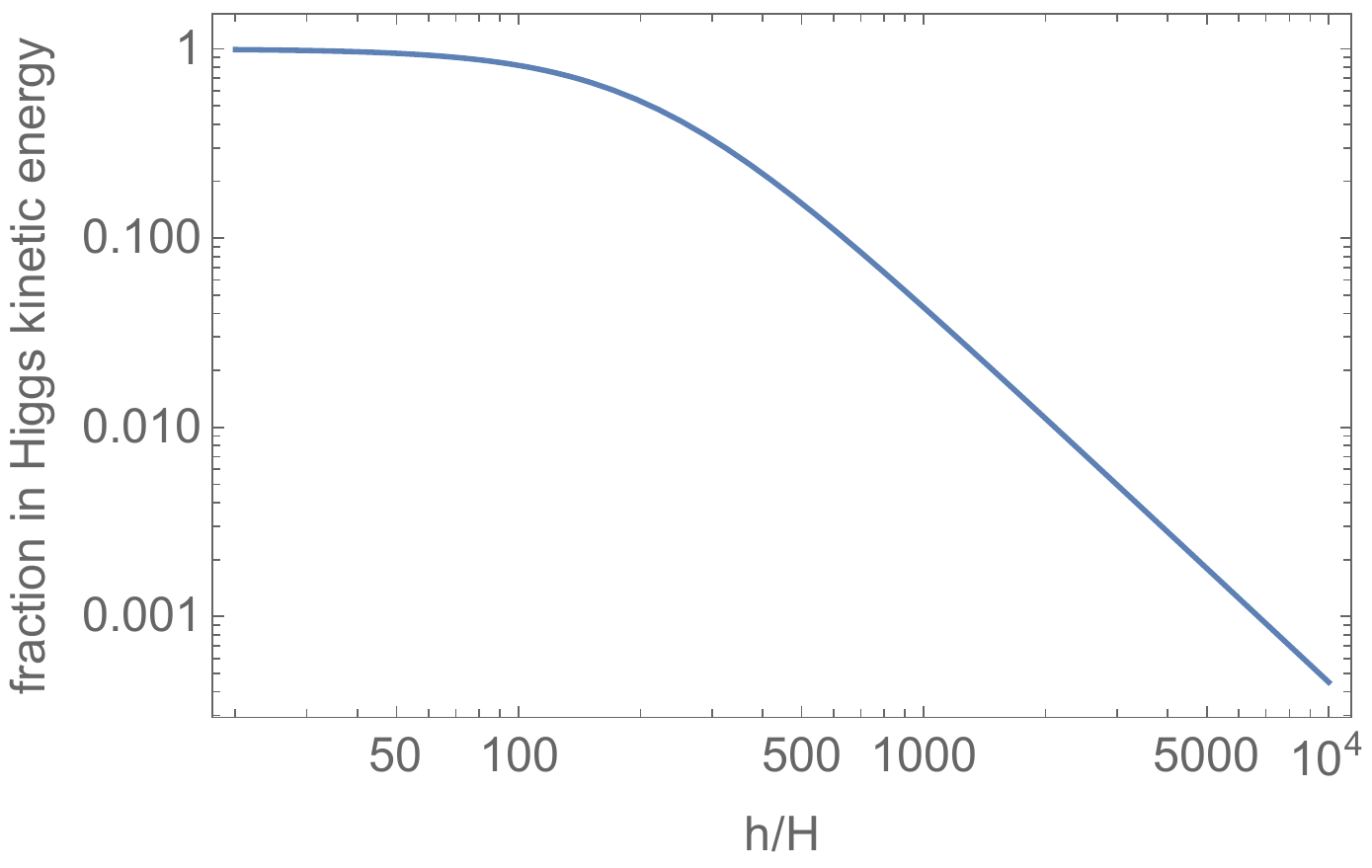}
  \end{center}\vspace{-4mm}
\caption{Fraction of Higgs potential energy converted to the kinetic energy of the rolling Higgs field, $\frac{1}{2}\dot{h}^2/{|V(h)|}$. The remainder goes into Higgs particle production.}
\label{fig:keratio}
\end{figure}

Beyond the slow-roll regime $h>h_{sr}$, Hubble friction is negligible, hence energy conservation dictates that the sum of the Higgs kinetic energy and the energy density in particles must be equal to the energy released from the Higgs potential, 
\be
\frac{1}{2}\dot{h}^2+\rho_P(h)\approx|V(h)|\,.
\ee
Substituting $\rho_P(h)$ from Eq.\,\ref{eq:rhoest} enables us to solve for $\dot{h}$ as a function of $h$,
\be
\dot{h}\sim \frac{\sqrt{\lambda/2}}{\sqrt{1+2\left(\frac{\lambda}{3}\right)^2\left(\frac{h}{H}\right)^2}} h^2\,.
\label{eq:newhdot}
\ee
The Higgs kinetic energy obtained from Eq.\,\ref{eq:newhdot} is shown in Fig.\,\ref{fig:keratio}. This shows that as the Higgs field rolls from $h_{sr}$ to $h_I$, an increasingly greater fraction of the released potential energy goes into particle production. At $h\gtrsim 10^3 H$, the Higgs kinetic energy is only a percent level component, and almost all of the released energy is instead in particles. 

We can also estimate the thermal correction to the Higgs potential due to the presence of the particle bath. Although the particle ensemble is produced in coherent states of various momenta and has not had the time to thermalize, we may nevertheless use energy conservation to assign an effective temperature $\Teff$, 
\be
\frac{\pi^2}{30} \Teff^4\approx \rho_P\approx |V(h)|\,~~\Rightarrow ~~\Teff\sim \left(\frac{15 \lambda}{2 \pi^2}\right)^{1/4}\!\!h\sim 0.3\, h\,.
\ee
The temperature correction to the Higgs potential from a thermal bath of Higgs particles at temperature $T$ is given by (see e.g.\,\cite{Kolb:1990vq,Jain:2019wxo})
\be
\Delta V_T=\frac{T^4}{2\pi^2}\int_0^\infty dz\,z^2\,\text{ln}\!\left(\frac{1-e^{-\sqrt{z^2+\frac{|m_h^2|}{T^2}}}}{1-e^{-z}}\right)\,.
\ee
With $T=\Teff$ above, it can be checked numerically that $\Delta V_T\!\sim\!0.03\, |V(h)|$, hence such thermal corrections do not appreciably modify the Higgs potential.\,\footnote{This conclusion remains unchanged even if the Higgs population decays into a thermal bath of SM particles.}

Let us briefly discuss the nature of the growing population of Higgs particles. Calculations of particle production from vacuum assume that the produced particle distribution is homogeneous and isotropic. However, it is also standard to interpret a high occupation number of scalars in a coherent state as a classical scalar wave with spatial extent $k^{-1}$ \cite{Felder:2000hj,Felder:2001kt} (however, see also \cite{Boucher:1988ua,Traschen:1990sw}), which therefore introduces sub-horizon sized spatial inhomogeneities in the otherwise homogeneous background field. The dynamics of such large inhomogeneities can form localized, pseudo-stable configurations known as oscillons, which remains a topic of active research \cite{Hertzberg:2010yz,Amin:2010dc,Amin:2010xe,Gleiser:2011xj,Amin:2011hj, Amin:2014eta}. In the most extreme cases, the development of large inhomogeneities can cause the fragmentation of the Higgs field itself\,\footnote{Similar scenarios have recently been encountered, for instance, in the context of resonant particle production leading to axion fragmentation \cite{Fonseca:2019ypl,Morgante:2021bks}.}, at which point it makes little sense to talk about a coherently evolving background Higgs field. Detailed understanding of such aspects requires lattice studies and is beyond the scope of this work; here we simply emphasize that understanding the dynamics of such configurations is necessary to understand the eventual fate of a spatial region with large Higgs inhomogeneities. 

Independent of such details, it is clear that the population of Higgs particles eventually becomes the dominant form of energy in the Universe as $h$  approaches $h_I$, and we expect this to terminate inflation locally (if inflation has not ended already due to inflaton dynamics) when $\rho_P\sim V_\phi+V(h)$.

\section{Post-Inflationary Evolution}
\label{sec:postinflation}

We now discuss post-inflationary evolution of regions where the Higgs has evolved to large field values $h > h_{sr}$, in particular whether such regions could be compatible with the existence of our Universe given the results in the previous sections. A definitive resolution of this question requires numerical simulations of the dynamics of the Higgs inhomogeneities as well as of spacetime itself, which is beyond the scope of this paper and will be addressed in future work \cite{futurework}; here we only provide some qualitative discussions.

 Higgs instability during preheating has been studied in previous works (such as \cite{Ema:2016kpf,Kohri:2016wof,Enqvist:2016mqj,Herranen:2015ima,Ema:2017rkk,Postma:2017hbk,Felder:2000hr,He:2020ivk,Kost:2021rbi}), but these focus on the Higgs in the EW minimum, studying the destabilizing effects of tachyonic instabilities, or resonant particle production through some inflaton-Higgs coupling. Here we are interested, instead, in configurations where the Higgs field is already in the unstable region, at $h > h_{sr}$. Previous works that considered this regime painted a bleak picture, concluding that the existence of such regions in our past lightcone has catastrophic consequences \cite{Kearney:2015vba,Espinosa:2015qea,East:2016anr,Strumia:2022kez}. Such conclusions were primarily based on two observations (see in particular \cite{East:2016anr} for detailed discussions): (1) any region with $h\gtrsim h_{sr}$ rapidly diverges to $h>h_I$ within a single e-fold, descending into crunching AdS, and post-inflationary reheating effects likely cannot provide large enough thermal corrections to restore the Higgs to the EW vacuum in such a short timeframe; (2) while the interior of the AdS region collapses into a black hole, the boundary of the AdS region grows \textit{outwards}, thereby engulfing other Hubble patches where the Higgs field might remain in the good (EW) vacuum. These results, however, were obtained within frameworks that only considered the interplay of the potential and kinetic energies of the Higgs and the inflaton, and must be reassessed in the presence of a large energy density in particles. 

Particle production improves the former consideration (1) in two respects: the Higgs velocity is slower in this regime (as seen from Eq.\,\ref{eq:newhdot} and Fig.\,\ref{fig:keratio}), and the local Hubble patch emerges out of inflation in a partially preheated state. One might hope that the slower Higgs velocity could significantly delay its approach to $h_I$; however, the time $\int_{h_{sr}}^{h_I}  dh/\dot{h}$  for the Higgs to roll from $h_{sr}$ to $h_I$ still evaluates (using Eq.\,\ref{eq:newhdot}) to roughly an e-fold. This is reasonable, as the friction from particle production becomes significant only towards the very late stages of this evolution. Nevertheless, the emergence out of inflation into a partially preheated state may aid with the preheating/reheating through the inflaton. In particular, if inflaton decay occurs via efficient resonant processes such as parametric resonance or tachyonic preheating, the coherent Higgs population (or its annihilation/decay products) already present can seed the growth of such resonances, making these even more efficient. Such explosive particle production from the inflaton can stabilize the Higgs potential through thermal corrections, driving it back to the EW vacuum. However, it should be noted that such effects require some nonminimal coupling between the inflaton and the SM, which will in general also modify the Higgs potential.

Such effects are unlikely to rescue Hubble patches where $h\sim h_I\sim 10^4\, H$, where $|V(h)|\sim V_\phi$, even with instantaneous reheating. On the other hand, in regions with the Higgs at somewhat smaller field values, e.g. $h\lesssim 10^3 H$, we have $|V(h)|/V_\phi\lesssim10^{-4}$, hence releasing even a small fraction of the inflaton energy density into particles could provide large enough corrections to stabilize the Higgs potential. It is well known that preheating from resonant effects can draw a significant fraction of the inflaton energy density within one, or a few, inflaton oscillations \cite{Felder:2000hj,Felder:2001kt,Boyanovsky:1996sq}. Fig.\,\ref{fig:keratio} shows that the kinetic energy of the Higgs is a percent level of the total energy released from the Higgs potential in this regime, hence the Higgs spends over an order magnitude longer time in this regime as a consequence of particle production, improving the prospects of rescuing such patches. Understanding whether this can in fact be realized requires model-dependent studies with specific models of inflation, as well as a more careful treatment of the particle population (including the evolution of inhomogeneities / oscillons discussed above), which is beyond the scope of this paper. Nevertheless, the above considerations at least make it more plausible that post-inflationary reheating can rescue regions with $h>h_{sr}$ from collapsing into AdS. 

Even the seemingly inevitable descent into AdS might not be catastrophic. The outward growth of these collapsing AdS regions, as found in \cite{East:2016anr} (see also \cite{DeLuca:2022cus,Strumia:2022kez}), was based on simulations that only considered potential and kinetic energies. It is possible that the inclusion of particles, which can become the dominant energy component in these patches, could change this outlook, and these AdS regions -- more accurately, regions with negative potential energy \cite{Felder:2002jk} but dominated by particle energy density -- could evolve, likely collapsing into black holes, without destroying neighboring regions where the Higgs is in the EW vacuum. A complete resolution of this question requires numerical simulations of the spacetime dynamics.

\section{Observable Signals}
\label{sec:observables}

Next, we briefly discuss some observable consequences of large excursions of the Higgs field to $h>h_{sr}$ in the early Universe. As demonstrated in the previous sections, such excursions are inevitably accompanied by a large energy density of Higgs particles in these Hubble patches. Note that while such particle densities could be a significant, even dominant, component of energy in the local patch, from a global viewpoint the Higgs field distribution is peaked around $h\sim 0$ \cite{Ellis:2009tp,Espinosa:2007qp,Elias-Miro:2011sqh,Kobakhidze:2013tn,Hook:2014uia,Kobakhidze:2014xda,Shkerin:2015exa,Kearney:2015vba,Espinosa:2015qea,East:2016anr,Grobov:2016llk,Kohri:2016qqv,Markkanen:2018pdo,DeLuca:2022cus}, and only a small fraction of inflating Hubble patches have the Higgs field at such large values. Therefore, at the end of inflation, this configuration results in very large inhomogeneities in an extremely small fraction of the Universe volume. Such inhomogeneities, if sufficiently numerous, can nevertheless leave several observable imprints in our Universe today. Again, the detailed nature of such signals can only be deduced after the post-inflationary dynamics of the Higgs inhomogeneities and spacetime is better understood, hence we only make some brief qualitative statements here. 

\textit{Imprints in the CMB:} Particle production during inflation is known to modify the primordial power spectrum
\cite{Chung:1999ve}, and produce CMB ``hotspots" \cite{Kim:2021ida}. 

\textit{Gravitational Waves:} Fluctuations in the Higgs field generated during inflation can source stochastic gravitational waves \cite{Espinosa:2018eve,Dufaux:2008dn,Cheong:2022gfc}, which can be observed with LISA, the Einstein Telescope, or Advanced-Ligo \cite{Espinosa:2018eve}. 

\textit{Black Holes:} The Higgs overdensities, if sufficiently large, can collapse into black holes, giving rise to a population of primordial black holes that could survive to present times \cite{Espinosa:2017sgp,Cheong:2022gfc} (and, with the right parameters, could even account for dark matter \cite{Espinosa:2017sgp}). 

Detailed studies of such signals in scenarios where the Higgs instability is accompanied by a large population of Higgs particles will be performed in future work \cite{futurework}.

\section{Discussion}
\label{sec:discussion}

This Letter has highlighted the importance of particle production from Higgs dynamics in its tachyonic phase during and after high scale inflation. It is shown that non-adiabatic evolution of the Higgs field beyond its slow-roll regime, $h>h_{sr}\sim 17 H$, excites Higgs particles out of the vacuum, and the tachyonic instability enhances these particle numbers exponentially, resulting in the released Higgs potential energy being converted almost entirely to particle energy density. Therefore, regions of space where the Standard Model Higgs field has fluctuated to such large values sustain a significant population of Higgs particles. Thus, any study of the Higgs field in this regime must take this sizeable particle density into account.

This remarkable result has many important implications. When such regions exit inflation, the emergent patch is already in a preheated state, which could facilitate efficient dissipation of the inflaton energy density if the inflaton decays through resonant effects that can build on this pre-existing particle abundance. This opens the prospects of rapid thermal corrections that could restore these regions to the electroweak vacuum instead of them evolving into anti-de Sitter space. Even if regions with large Higgs field values fall into AdS, it is not clear whether, in the presence of large particle densities, such occurrences are necessarily catastrophic for the macroscopic Universe. Likewise, the presence of such large inhomogeneities in parts of the post-inflationary Universe can also leave observable imprints in the CMB, gravitational waves, or primordial black holes, which would represent tantalizing evidence of the realization of the tachyonic phase of the SM Higgs in our cosmological history. 

The results in this Letter could also hold relevance for questions of a more fundamental nature.  The Higgs potential is metastable only for a very narrow range of parameters (Higgs and top quark masses), hence its realization in nature is somewhat of a mystery. In this context, it is possible that the metastable vacuum exists because it serves some important purpose in the early Universe, which could be related to its tachyonic phase, and perhaps the explosive particle production that comes with it. The tachyonic regime also plays an intriguing role in the context of eternal inflation: {any} patch that inflates for too long will inevitably find itself in the tachyonic phase of the Higgs, which will trigger a rapid growth of inhomogeneities, terminating inflation locally; in this sense, the tachyonic Higgs could act as a regulator of eternal inflation.\,\footnote{See also \cite{Jain:2019gsq,Jain:2019wxo} for related ideas.}

Several aspects and implications of Higgs particle production were only touched upon briefly and qualitatively here, and require further detailed study. Of paramount importance is a better understanding of the evolution of the large particle number densities or inhomogeneities, which requires lattice studies. It would also be insightful to study the post-inflationary evolution of patches with large Higgs number densities within specific models of inflation and (p)reheating, to understand the extent to which large Higgs field values can be saved from collapsing into AdS. Numerical simulations of the negative potential energy regimes with a significant particle energy density are needed to clarify whether such regions can be compatible with the existence of our Universe. Likewise, careful derivation of the nature of observable signals of the tachyonic phase and accompanying particle production, in particular in the CMB, gravitational waves, and primordial black holes, will be crucial in making more direct connections with ongoing and future experimental programs. These aspects will be addressed in greater detail in future work \cite{futurework}.

%%%%%%%%%%%%%%%%%%%%%%%%%%%%%%%%%%%%%%%%%%%%%%%%%%
\section*{Acknowledgments} 
%%%%%%%%%%%%%%%%%%%%%%%%%%%%%%%%%%%%%%%%%%%%%%%%%%

It is a pleasure to thank Gian Giudice for multiple insightful discussions at various stages of this project. The author is also grateful to Antonio Riotto, Geraldine Servant, Alexander Westphal, and Kathryn Zurek for helpful discussions and feedback, and thanks the Berkeley Center for Theoretical Physics and the Lawrence Berkeley
National Laboratory for hospitality during the completion of this project. This work is supported by the Deutsche Forschungsgemeinschaft under Germany's Excellence Strategy - EXC 2121 Quantum Universe - 390833306. 

\bibliography{tachyonic_higgs}{}

\end{document}